\documentclass[a4paper,10pt,]{article}
\usepackage[UKenglish]{babel}
\usepackage[utf8]{inputenc}
\usepackage[T1]{fontenc}

\usepackage[amssymb]{SIunits}
\usepackage[font=small]{caption}
\usepackage[nameinlink,capitalize]{cleveref}
\usepackage[top=12mm,bottom=12mm,left=30mm,right=30mm,head=12mm,includeheadfoot]{geometry}
\usepackage{amsmath}
\usepackage{amssymb}
\usepackage{amsthm}
\usepackage{authblk}
\usepackage{braket}
\usepackage{cancel}
\usepackage{colonequals}
\usepackage{esvect}
\usepackage{graphicx}
\usepackage{hyperref}
\usepackage{nicefrac}
\usepackage{physics}
\usepackage{xspace}
\usepackage[numbers,sort&compress]{natbib}

\hypersetup{pdfauthor={Claudius Hubig},pdftitle={Use and implementation of autodifferentiation in tensor network methods with complex scalars}}

\newcommand{\stensor}{\texttt{STensor}\xspace}
\newcommand{\unl}[1]{\underline{#1}}
\newcommand{\inv}[1]{ {{#1}^{-1}} }
\newcommand{\xp}[0]{{x^\prime}}

\begin{document}

\title{Use and implementation of autodifferentiation in tensor network methods with complex scalars}
\author{Claudius Hubig}
\affil{\small{Max-Planck-Institut für Quantenoptik, Hans-Kopfermann-Strasse 1, 85748 Garching}}
\date{\today}

\maketitle

\begin{abstract}
  Following the recent preprints
  \href{https://arxiv.org/abs/1903.09650}{arXiv:1903.09650} and
  \href{https://arxiv.org/abs/1906.04654}{arXiv:1906.04654} we comment
  on the feasibility of implementation of autodifferentiation in
  standard tensor network toolkits by briefly walking through the
  steps to do so. The total implementation effort comes down to less
  than 1000 lines of additional code.

  We furthermore summarise the current status when the method is
  applied to cases where the underlying scalars are complex, not real
  and the final result is a real-valued scalar. It is straightforward
  to generalise most operations (addition, tensor products and also
  the QR decomposition) to this case and after the initial submission
  of these notes, also the adjoint of the complex SVD has been found.
\end{abstract}

\tableofcontents

\section{Introduction}
\label{sec:intro}

Autodifferentiation (AD) is a technique to automatically compute the
gradient of a computer program by defining the partial derivatives of
the individual building blocks of the program and then using the chain
rule to combine the partial derivatives of the steps run during the
computer program execution into an overall derivative. The method is a
standard tool in the field of machine learning, where automatically
computed gradients are used for the optimisation of neural networks
(called ``backpropagation''). It was recently by Liao et
al\cite{liao19:_differ_progr_tensor_networ} into the field of tensor
network methods in the context of infinite projected entangled pair
states and has already been applied to other problems as
well\cite{torlai19:_wavef}.

Here, we want to firstly demonstrate that the implementation of
reverse-mode autodifferentiation is straightforward in a standard
tensor networks toolkit as typically used in the community without
relying on existing toolchains such as TensorFlow or PyTorch. Such a
``native'' implementation has the advantage of exploiting all the
standard tricks to speed up condensed-matter simulations, in
particular the use of symmetries\cite{mcculloch02:_abelian,
  mcculloch02:_collec_phenom_stron_correl_elect_system,
  mcculloch07:_from, singh10:_tensor, singh11:_tensor_u,
  singh12:_tensor_su, weichselbaum12:_non,
  hubig17:_symmet_protec_tensor_networ, hubig18:_abelian} to enforce a
block structure on tensors.

Secondly, we want to summarise the current status when the scalars
employed in the tensor network intermittently are not real but complex
(e.g. \texttt{std::complex<double>} instead of \texttt{double}), with
only the final result, such as an energy or cost function, being
real-valued again.

\section{Implementation Effort for Reverse-Mode AD}
\label{sec:impleffort}

While the authors have implemented reverse-mode AD in the
\textsc{SyTen}\cite{hubig:_syten_toolk} tensor networks toolkit, the
implementation can likely proceed in much the same way in any other
codebase. The \stensor class introduced into the
\textsc{SyTen} toolkit in 2018 provides named tensor indices,
automatic tensor products over equal-named indices, automatic handling
of fermionic indices without the need for swap
gates\cite{barthel09:_contr, bultinck17:_fermion} and can be combined
with an \texttt{AsyncCached<>} template to transparently and
asynchronously cache tensor contents to disk when not needed.

Implementation of autodifferentiation support proceeded in two steps:

\subsection{\texttt{ComputeNode} history storage}

The \stensor class was extended by two objects: an autodifferentiation
ID, uniquely identifying the specific tensor object, and a shared
pointer to a \texttt{ComputeNode} object. The \texttt{ComputeNode}
class represents a particular step of a calculation, e.g. an addition
or tensor-tensor product. Each \texttt{ComputeNode} stores a list of
shared pointers to the compute nodes associated with the input tensors
of the operation, a list of IDs of the output tensors generated
(typically 1, but e.g. the QR decomposition produces two output
tensors), an adjoint evaluation function and cached copies of the
tensors necessary for the adjoint evaluation function to run. It was
useful to also store empty tensors of the shape of the output tensors
(and hence output adjoints) to more generally handle cases where these
adjoints may be zero due to the result not depending on that
particular output.

When enabling autodifferentiation on a specific tensor, an initial
compute node is created for this tensor. Tensor operations create
subsequent compute nodes which build up a directed acyclic graph.

Once the desired result value is obtained, requesting the
autodifferenation of this output value with respect to a valid
autodifferentiation ID causes the graph to be traversed backwards
until the compute node producing this ID is found. During this
traversal, the graph is double-linked such that each node knows which
other nodes rely on its output as an input. When the original node is
found, its output adjoint with respect to the result value
autodifferentiation ID is requested. This request percolates down the
tree, each node first computing the adjoints of its outputs by
requesting the adjoints of the inputs from its downstream nodes and
those adjoints of inputs being evaluated by the associated adjoint
evaluation function stored in each node.

For efficiency, the result of the output adjoint evaluation is cached
in each node such that multiple differentials can be computed
easily. For convenience and debugging, it is also possible to draw the
directed acyclic graph representing the computation (in our case, by
producing an input file to the `dot` program which handled the actual
drawing). The overall framework including the definition of compute
nodes, their clean-up and the calculation of adjoints (given the
adjoint evaluation functions defined elsewhere) can be done in about
350 physical lines of C++.

\begin{figure}
  \centering
  \includegraphics[width=0.8\textwidth]{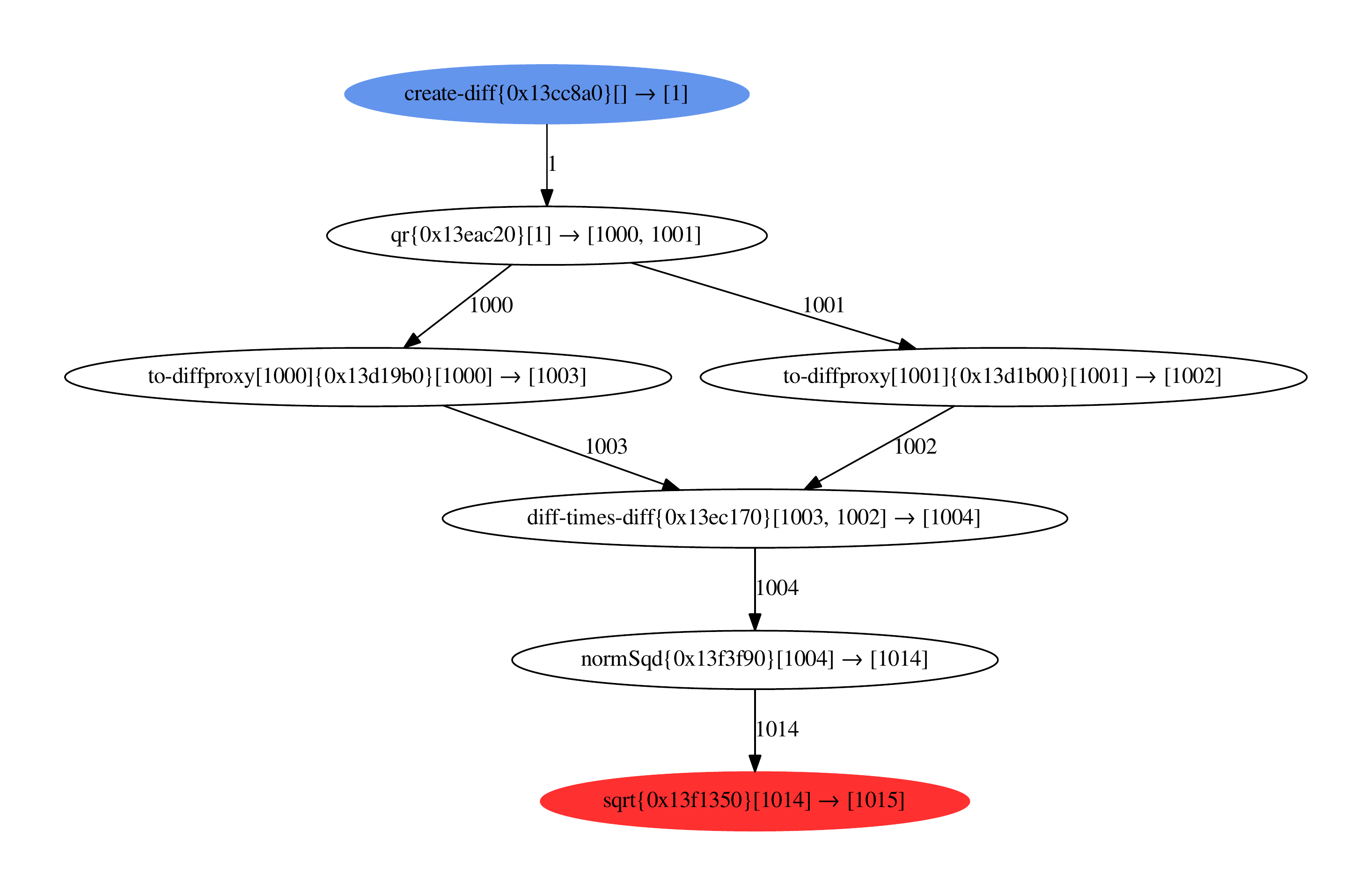}
  \caption{An exemplary computation tree resulting from the evaluation
    $ Q, R = \mathrm{qr}(X); Y = Q \cdot R; Z = \sqrt{ || Y
      ||^2_{\mathrm{Frob}} } $.
    The tensor-tensor product takes two temporary proxy objects, here
    showing up as \texttt{to-diffproxy} operations.}
  \label{fig:example-tree}
\end{figure}

\subsection{Definition of \texttt{AdjointEvaluator} functions}

Once the basic framework exists, one has to adapt every function
manipulating an \stensor object to potentially store its computation
and adjoint evaluation function in a compute node, if that \stensor
object has autodifferentiation enabled. If one first changes every
function to assert that its input tensors do not have
autodifferentiation enabled the additional definitions can be added
one by one without fear of introducing undetected errors.

In the following, we will consider some examples of such adjoint
evaluation functions. These function should, when called from a
compute node which has as input a tensor $I$ and produced a tensor $T$
with final scalar result $R$, evaluate the ``input adjoint''
$\frac{\partial R}{\partial I}$ where they can rely on
$\frac{\partial R}{\partial T}$ being available as the ``output
adjoint'' sum of the input adjoints of downstream nodes.

As a first example, consider a function which changes the name of one
of the tensor legs, that is,
$x_{\ldots, i, \ldots} \to f(x_{\ldots,i,\ldots} =
x_{\ldots,j,\ldots}$.
Working reverse, we need to change the tensor leg $j$ of the output
adjoint back to have the name $i$. If the output adjoint has such a
leg, this can be done simply by renaming. If it does not (typically
not the case if the final result which we differentiate is a scalar),
we need to take the outer product of the output adjoint and an
identity tensor mapping $i$ to $j$.

Second, consider the case of a tensor-tensor addition $I_1 + I_2 = T$.
The partial derivative $\partial T/\partial I_{1,2}$ is one, hence
when the input adjoint with respect to either $I_1$ or $I_2$ is
requested, we can simply return the output adjoint
$\partial R/\partial T$. Note that neither of these two cases require
storing either the input or output tensors of the node.

Third, consider the case of hermitiation conjugation, that is
$X \to f(X) = X^\dagger$. For complex-valued scalars, it is easiest to
assume that $\partial X^\dagger/\partial X$ is zero and hence simply
discard all earlier history and consider the resulting tensor
$X^\dagger$ as a potential new origin tensor. In the real-valued case,
the situation is more complicated as $X^T$ does depend on $X$ and the
adjoint evaluator has to return the transpose of the output adjoint it
obtains from downstream. This can be seen by writing the transposition
as a series of tensor-tensor
products\cite{laue18:_comput_higher_order_deriv_matrix_tensor_expres}
with tensors of the form $\delta^{ii}$ or $\delta_{jj}$ to exchange
upstairs and downstairs indices. The local partial derivative is then
just those tensors, multiplying them into the output adjoint
transposes it.

Fourth, the case of tensor-tensor products $I_1 \cdot I_2 = T$ can
also be handled straightforwardly, however, now it is necessary to
store both input tensors $I_{1,2}$. The adjoint evaluation simply has
to multiply the downstream adjoint by either $I_1$ or $I_2$.

\subsubsection{Tensor decompositions}

For both the eigenvalue decomposition, the QR decomposition and the
singular value decompositions, expressions for the adjoints in the
real case are straightforwardly available\cite{walter10:_higher_qr,
  seeger17:_auto_differ_linear_algeb,
  townsend16:_differ_singul_value_decom,
  liao19:_differ_progr_tensor_networ}. While implementing those
operations, in particular the element-wise Hadamard products, is
tricky, this can also be done. Note that the adjoints of these
decompositions typically rely on matrix inversion (either of the
singular value matrix, the eigenvalue matrix or the upper-triangular
matrix in the QR), which may cause numerical problems unless
stabilised\cite{liao19:_differ_progr_tensor_networ}.

\section{Complex Scalars}

The standard machine learning toolkits mostly handle the case of
real-valued scalars. However, in quantum physics, complex scalars are
not often avoidable. It is hence interesting to see how reverse-mode
autodifferentiation applies to the case of complex scalars if the
final result value $r$ is limited to the reals: That is, we consider
tensor operations that take complex-valued input tensors and finally
produce a real value, $f: \mathbb{C}^n \to \mathbb{R}$. This is for
example the case where we wish to obtain the gradient of a physical
observable (such as the energy) or if we otherwise want to optimise
some cost function (e.g. entanglement over a bond). Naturally, the
functions we consider are typically not analytic, i.e. depend on both
the input variables $z$ and their complex conjugates $z^\dagger$. In
those cases, it is most natural to consider $z$ and $z^\dagger$ to be
independent variables and make use of the Wirtinger derivative
\begin{equation}
  \frac{\partial}{\partial z} \equiv \frac{1}{2} \left[ \frac{\partial}{\partial \mathrm{Re}(z)} - \mathrm{i} \frac{\partial}{\partial \mathrm{Im}(z)} \right] \;.
\end{equation}
Evaluating $\frac{\partial r(Z,Z^\dagger)}{\partial Z}$ for a complex
tensor $Z$ and a real-valued scalar $r$ yields the gradient of
$r(Z,Z^\dagger)$ at position $Z$, its hermitian conjugate
$\left(\frac{\partial r(Z,Z^\dagger)}{\partial Z}\right)^\dagger$ is
the conjugate gradient (which has the same dimensions as $Z$) and may
be used to take a step in the direction of steepest
ascent/descent. Note that
\begin{equation}
  \left(\frac{\partial r(Z,Z^\dagger)}{\partial Z}\right)^\dagger = \frac{\partial r(Z,Z^\dagger)}{\partial Z^\dagger} \;.
\end{equation}
See \cite{hunger07:_introd_compl_differ_compl_differ} for a
pedagogical overview and further details.

\subsection{Standard Operations}

All standard operations such as addition, tensor products or index
renaming translate straightforwardly from the real to the complex case
and no special handling in code is required. Only taking the hermitian
conjugate requires differentiating between the real and complex case
-- in the former, it is equivalent to a simple tensor transpose (which
is differentiable), in the latter, it creates a new independent
variable.

Functions which produce either the real or imaginary part of their
inputs are best represented as additions/subtractions with the complex
conjugate followed by multiplication by a scalar $\frac{1}{2}$ or
$\frac{-\mathrm{i}}{2}$. The complex conjugate is an independent
variable and hence does not matter, the multiplication by a scalar
simply needs to be translated to the downstream adjoint.

\subsection{QR Decomposition}

While the available literature only discusses real QR decompositions,
the complex QR decomposition is also unique if one requires the
diagonal of $R$ to be real and positive. It is useful to insert a
manual check to this end in the QR decomposition, depending on the
underlying library/LAPACK implementation.

Thankfully, the calculation steps done to produce adjoints of $Q$ and
$R$ in the real case translate straightforwardly to the complex case:

Let in the following $A_i^j$ denote elements of a matrix
$A \in \mathbb{C}^{\dim(i) \times \dim(j)}$ with
$\dim(i) \geq \dim(j)$ (``tall''). The element-wise complex conjugate
of a matrix $X_i^j$ is given by $\unl{X}_i^j$. The QR decomposition of
$A$ is given by matrices $Q \in \mathbb{C}^{\dim(i) \times \dim(x)}$
and $R \in \mathbb{C}^{\dim(x) \times \dim(j)}$ such that
\begin{align}
A_a^i & = Q_a^x R_x^i \label{eq:aqr} \;.
\end{align}
Furthermore $Q^\dagger Q = \mathbf{1}$, i.e.:
\begin{equation}
\unl{Q}^x_a Q_a^{\xp} \delta^{a a} = \delta^{x \xp} \;. \label{eq:qqi}
\end{equation}
Furthermore, we have projectors $L_\alpha^\beta = 1 \textrm{ iff } \alpha > \beta$, $U_\alpha^\beta = 1 \textrm{ iff } \beta > \alpha$ and
$D_\alpha^\beta = 1 \textrm{ iff } \alpha \equiv \beta$. With those projectors, the triangularity of $R$ is
\begin{equation}
  R_x^i L_x^i = 0_x^i\;.
\end{equation}

\subsubsection{Aim}

The total differential of a real-valued function $\phi$ which we are interested in is
(cf.~\cite{hunger07:_introd_compl_differ_compl_differ}, pg.~11)
\begin{equation}
  \dd \phi = \frac{\partial \phi}{\partial A_a^i} \dd A_a^i + \frac{\partial \phi}{\partial \unl{A}_a^i} \unl{\dd A}_a^i = a^a_i \dd A_a^i + \unl{a}^a_i \unl{\dd A}_a^i
\end{equation}
with the adjoint $a$ the prefactor of $\dd A$ in $\dd \phi$, of which we want to take twice the real part. Because the
adjoint of $\unl{A}$ is equal to the complex-conjugate of the adjoint of $A$ for real-valued $\phi$, we can choose to
either evaluate $a$ and complex-conjugate it or evaluate $\unl{a}$ directly. In any case, $\unl{\dd A}$ is only the
complex conjugate of $\dd A$.

Written in terms of $Q$ and $R$, this is
\begin{equation}
  \dd \phi = q_x^a \dd Q_a^x + r_i^x \dd R_x^i + \unl{q}_x^a \unl{\dd Q}_a^x + \unl{r}_i^x \unl{\dd R}_x^i \quad. \label{eq:dphi}
\end{equation}
Our task then is to find expressions for the differentials $\dd Q$ etc to re-express those as differentials of $\dd A$
and $\unl{\dd A}$. Subsequently, the desired adjoint $a$ will simply be the prefactor of $\dd A$.

\subsubsection{Expression for $\dd Q_a^x$}

From \eqref{eq:aqr}, we have
\begin{align}
  \dd A_a^i & = \dd Q_a^x R_x^i + Q_a^x \dd R_x^i \\
  \dd Q_a^x R_x^i & = \dd A_a^i - Q_a^x \dd R_x^i \\
  \dd Q_a^x R_x^i \inv{R}^y_i & = \dd A_a^i \inv{R}^y_i - Q_a^x \dd R_x^i \inv{R}^y_i \\
  \dd Q_a^y & = \dd A_a^i \inv{R}^y_i - Q_a^x \dd R_x^i \inv{R}^y_i \;. \label{eq:dq}
\end{align}

\subsubsection{Expression for $\dd R_x^i$}

Differentiating \eqref{eq:qqi} gives
\begin{equation}
\unl{\dd Q}^x_a Q_a^{\xp} \delta^{a a} + \unl{Q}^x_a \dd Q_a^{\xp} \delta^{a a} = \dd \delta^{x \xp} = 0\;.
\end{equation}
and hence
\begin{equation}
  Q_a^{\xp} \unl{\dd Q}^x_a \delta^{a a} = - \unl{Q}^x_a \dd Q_a^{\xp} \delta^{a a}
\end{equation}
LHS and RHS are equal under simultaneous exchange of
$x \leftrightarrow \xp$, complex conjugation and multiplication by
$-1$, the matrix in $x, \xp$ is antihermitian. Then multiplying
\eqref{eq:dq} by $\unl{Q}^x_a \delta^{aa}$ gives:
\begin{align}
  \unl{Q}^x_a \delta^{aa} \dd Q_a^y & = \unl{Q}^x_a \delta^{aa} \dd A_a^i \inv{R}^y_i - \unl{Q}^x_a \delta^{aa} Q_a^z \dd R_z^i \inv{R}^y_i \\
                                    & = \unl{Q}^x_a \delta^{aa} \dd A_a^i \inv{R}^y_i - \delta^{xz} \dd R_z^i \inv{R}^y_i \;.
\end{align}
Due to the antihermiticity of the LHS, the same must hold for the RHS
under exchange $x \leftrightarrow y$ and complex conjugation:
\begin{align}
  \unl{Q}^x_a \delta^{aa} \dd A_a^i \inv{R}^y_i - \delta^{xz} \dd R_z^i \inv{R}^y_i = - Q^y_a \delta^{aa} \unl{\dd A}_a^i \unl{\inv{R}}^x_i + \delta^{yz} \unl{\dd R_z^i} \unl{\inv{R}^x_i} \;.
\end{align}
Sorting the terms, we get
\begin{align}
  \unl{Q}^x_a \delta^{aa} \dd A_a^i \inv{R}^y_i + Q^y_a \delta^{aa} \unl{\dd A}_a^i \unl{\inv{R}}^x_i = \delta^{yz} \unl{\dd R_z^i} \unl{\inv{R}^x_i} + \delta^{xz} \dd R_z^i \inv{R}^y_i
\end{align}
On the right-hand side, $\delta^{xz}\dd R_z^i \inv{R}^y_i$ is
upper-triangular in $(x,y)$, whereas the other term there is its
complex-conjugate. Assuming that $\dd R$ and $\inv{R}$ can be chosen
to be real on the diagonal, element-wise multiplication by
$E^{xy} = 2 D^{xy} + U^{xy}$ simply obtains twice the second summand
on the RHS and allows solving for $\dd R$:
\begin{align}
  \frac{1}{2} E^{xy} \left( \unl{Q}^x_a \delta^{aa} \dd A_a^i \inv{R}^y_i + Q^y_a \delta^{aa} \unl{\dd A}_a^i \unl{\inv{R}}^x_i \right) & = \delta^{xz} \dd R_z^i \inv{R}^y_i \\
  \frac{1}{2} E^{xy} \left( \unl{Q}^x_a \delta^{aa} \dd A_a^i \inv{R}^y_i + Q^y_a \delta^{aa} \unl{\dd A}_a^i \unl{\inv{R}}^x_i \right) R_y^j & = \delta^{xz} \dd R_z^i \inv{R}^y_i R_y^j \\
  \frac{1}{2} E^{xy} \left( \unl{Q}^x_a \delta^{aa} \dd A_a^i \inv{R}^y_i + Q^y_a \delta^{aa} \unl{\dd A}_a^i \unl{\inv{R}}^x_i \right) R_y^j & = \delta^{xz} \dd R_z^j  \\
  \frac{1}{2} E^{xy} \left( \unl{Q}^x_a \delta^{aa} \dd A_a^i \inv{R}^y_i + Q^y_a \delta^{aa} \unl{\dd A}_a^i \unl{\inv{R}}^x_i \right) R_y^j \delta_{xz} & = \dd R_z^j
\end{align}

\subsubsection{The Total Differential}
Returning to \eqref{eq:dphi}, we have:
\begin{align}
  \dd \phi & = q_x^a \dd Q_a^x + r_i^x \dd R_x^i + \unl{q}_x^a \unl{\dd Q}_a^x + \unl{r}_i^x \unl{\dd R}_x^i \\
           & = q_x^a \left( \dd A_a^i \inv{R}^x_i - Q_a^z \dd R_z^j \inv{R}^x_j \right) + r_j^z \dd R_z^j + \unl{q}_x^a \left( \unl{\dd A_a^i} \unl{\inv{R}}^x_i - \unl{Q}_a^z \unl{\dd R}_z^j \unl{\inv{R}}^x_j \right) + \unl{r}_j^z \unl{\dd R}_z^j \\
           & = q_x^a \dd A_a^i \inv{R}^x_i + \unl{q}_x^a \unl{\dd A_a^i} \unl{\inv{R}}^x_i  + \left( -q_x^a Q_a^z \inv{R}^x_j + r_j^z \right)\dd R_z^j + \left( -\unl{q}_x^a \unl{Q}_a^z \unl{\inv{R}}^x_j + \unl{r}_j^z \right) \unl{\dd R_z^j} \\
           & = q_x^a \dd A_a^i \inv{R}^x_i + \unl{q}_x^a \unl{\dd A_a^i} \unl{\inv{R}}^x_i \\
           & + \left( -q_u^b Q_b^z \inv{R}^u_j + r_j^z \right) \frac{1}{2} E^{xy} \left( \unl{Q}^x_a \delta^{aa} \dd A_a^i \inv{R}^y_i + Q^y_a \delta^{aa} \unl{\dd A}_a^i \unl{\inv{R}}^x_i \right) R_y^j \delta_{xz} \\
           & + \left( -\unl{q}_u^b \unl{Q}_b^z \unl{\inv{R}}^u_j + \unl{r}_j^z \right) \frac{1}{2} E^{xy} \left( Q^x_a \delta^{aa} \unl{\dd A}_a^i \unl{\inv{R}}^y_i + \unl{Q}^y_a \delta^{aa} \dd A_a^i \inv{R}^x_i \right) \unl{R}_y^j \delta_{xz}\;.
\end{align}
The coefficient $a^a_i$ of $\dd A_a^i$ is then:
\begin{align}
  a^a_i & = q_x^a \inv{R}^x_i \\
        & + \frac{1}{2} \left( -q_u^b Q_a^z \inv{R}^u_j + r_j^z \right) E^{xy} \unl{Q}^x_a \delta^{aa} \inv{R}^y_i R_y^j \delta_{xz} \\
        & + \frac{1}{2} \left( -\unl{q}_u^b \unl{Q}_b^z \unl{\inv{R}}^u_j + \unl{r}_j^z \right) E^{xy} \unl{Q}^y_a \delta^{aa} \inv{R}^x_i \unl{R}_y^j \delta_{xz} \\
        & = q_x^a \inv{R}^x_i \\
        & + \frac{1}{2} \left( -q_u^b Q_b^z \inv{R}^u_j R_y^j + r_j^z R_y^j \right) E^{xy} \unl{Q}^x_a \delta^{aa} \inv{R}^y_i \delta_{xz} \\
        & + \frac{1}{2} \left( -\unl{q}_u^b \unl{Q}_b^z \unl{\inv{R}}^u_j \unl{R}_y^j + \unl{r}_j^z  \unl{R}_y^j \right) E^{xy} \unl{Q}^y_a \delta^{aa} \inv{R}^x_i \delta_{xz} \\
        & = q_x^a \inv{R}^x_i \\
        & + \frac{1}{2} \left( -q_y^b Q_b^z  + r_j^z R_y^j \right) E^{xy} \unl{Q}^x_a \delta^{aa} \inv{R}^y_i \delta_{xz} + \frac{1}{2} \left( -\unl{q}_y^b \unl{Q}_b^z + \unl{r}_j^z  \unl{R}_y^j \right) E^{xy} \unl{Q}^y_a \delta^{aa} \inv{R}^x_i \delta_{xz} \\
        & = q_x^a \inv{R}^x_i  \qquad \textrm{ now rename $x \leftrightarrow y$ in second term} \\
        & + \frac{1}{2} \left( -q_y^b Q_b^z  + r_j^z R_y^j \right) E^{xy} \unl{Q}^x_a \delta^{aa} \inv{R}^y_i \delta_{xz} + \frac{1}{2} \left( -\unl{q}_x^b \unl{Q}_b^z + \unl{r}_j^z  \unl{R}_x^j \right) E^{yx} \unl{Q}^x_a \delta^{aa} \inv{R}^y_i \delta_{yz} \\
        & = q_x^a \inv{R}^x_i + \frac{1}{2} \left( -q_y^b Q_{bx}  + r_{jx} R_y^j \right) E_{xy} \unl{Q}^x_a \delta^{aa} \inv{R}^y_i + \frac{1}{2} \left( -\unl{q}_x^b \unl{Q}_{by} + \unl{r}_{jy}  \unl{R}_x^j \right) E_{yx} \unl{Q}^x_a \delta^{aa} \inv{R}^y_i \\
        & = q_x^a \inv{R}^x_i + \frac{1}{2} \unl{Q}^x_a \delta^{aa} \inv{R}^y_i \left[ \left( r_{jx} R_y^j - q_y^b Q_{bx} \right) E_{xy} + \left( \unl{r}_{jy}  \unl{R}_x^j - \unl{q}_x^b \unl{Q}_{by} \right) E_{yx} \right] \\
        & = q_x^a \inv{R}^x_i + \unl{Q}^x_a \delta^{aa} \inv{R}^y_i \mathrm{copyhermupper2lower_{rows: x, cols: y}}\left[ r_{jx} R_y^j - q_y^b Q_{bx} \right] \\
        & = q_y^a \inv{R}^y_i + \unl{Q}^x_a \delta^{aa} \inv{R}^y_i \mathrm{copyhermupper2lower_{rows: x, cols: y}}\left[ r_{jx} R_y^j - q_y^b Q_{bx} \right] \\
        & = \inv{R}^y_i \left( q_y^a + \unl{Q}^{xa} \mathrm{copyhermupper2lower_{rows: x, cols: y}}\left[ r_{jx} R_y^j - q_y^b Q_{bx} \right] \right)
\end{align}
Apart from the complex conjugation of $Q^{xa}$ which is missing in the
real case, this is the same expression as derived elsewhere.

\subsection{The Singular Value Decomposition}

The singular value decomposition $X = U \cdot S \cdot V$ in the real
case is only unique up to factors of $\pm 1$ in columns of $U$ and
rows of $V$. This not-uniqueness is not differentiable, however, given
a specific $X$ and a deterministic singular value decomposition
routine, we will for each choice only obtain either $+1$ or $-1$ and
the automatically computed gradient is not affected by this. In the
complex case, however, this prefactor can be any complex phase
$e^{i \phi}$ with $\phi$ a continuous variable. This additional
freedom is potentially differentiable. It appears easiest to gauge it
away by requiring that (e.g.) the first column of $V$ has to be real
and positive.

Nevertheless, doing the same derivation as in the real case leads to
problems: all of the published results rely on the antihermiticity of
$U^\dagger dU$ (obtained from
$U^\dagger U = 1 \Rightarrow \mathrm{d}U^\dagger U = -U^\dagger
\mathrm{d}U$)
to sum up two parts, in doing so, however, the potentially imaginary
diagonals of those parts are lost (in the real case, the diagonal is
zero). One possible solution may be to use the gauge freedom above to
enforce that this diagonal is zero.

This problem only affects the adjoints with respect to $U$ and $V$. If
the cost function only depends on the singular values $S$, the result
from the real case carries over directly as
$\mathrm{adj}(x) = U^\dagger \mathrm{adj}(s) V^\dagger$.

Solving this problem would allow for the direct use of
autodifferentiation also in cases where scalars have to be
complex-valued and operations include a SVD to e.g. truncate and
project tensor legs after a renormalisation step.

\subsubsection{Update}

After the first write-up of these notes, the proper adjoint of the
complex SVD has been found\cite{wan19:_autom_differ_compl_valued_svd}.

On this topic, it is useful to stress that the cost function
differentiated with respect to the input matrix of the SVD should be
gauge-invariant. That is, given $M = U S V^\dagger$, the
cost function $f(U, S, V) = f(M)$ must only depend on $U$ and $V$ in
such a way that it is invariant under insertion of complex phase
factor matrices $\Lambda$ into the SVD. Given two singular
value decompositions:
\begin{align}
  M & = U S V \\
  M & = U \Lambda \Lambda^{-1} S \Lambda \Lambda^{-1} V \\
  \Leftrightarrow M & = \tilde{U} S \tilde{V}
\end{align}
any differentiable cost function $f(U, S, V)$ should be constant under
the replacement $U \to \tilde{U}, V \to \tilde{V}$. For example, the
cost function $f = \mathfrak{Re}(\mathrm{tr}(U))$ is not
gauge-invariant while the functions
$f = \mathfrak{Re}(\mathrm{tr}(U V))$ and
$f = \mathfrak{Re}(\mathrm{tr}(U U^\dagger))$ are
gauge-invariant.

\section{Testing}

Testing the implementation is easiest if the final output is a
real-valued scalar, in the complex case, this is likely the only
sensible choice. To obtain this scalar, one may either take norms of
result tensors, scalar products of result tensors with fixed
predefined random tensors or select a particular tensor element by
taking the scalar product with a selection tensor (which is zero in
all but one entry). The complex-valued scalar may be translated into a
real-valued one by taking only its real or imaginary part, this
operation is also differentiable.

Once a function $f(Z)$ containing our test candidate and producing a
real-valued scalar has been obtained, we generate a small perturbation
$\epsilon$ of the same shape as $Z$. The difference
$f(Z+\epsilon) - f(Z)$ can then be compared to the scalar product
$\epsilon \cdot \partial f(Z)/\partial Z$.

\section*{Acknowledgements}
This work was funded through ERC Grant QUENOCOBA, ERC-2016-ADG (Grant
no. 742102). Helpful discussions with P.~Emonts and L.~Hackl are
gratefully acknowledged.

\end{document}